\documentclass[pdflatex,sn-mathphys-num]{sn-jnl}

\usepackage{graphicx}%
\usepackage{multirow}%
\usepackage{amsmath,amssymb,amsfonts}%
\usepackage{amsthm}%
\usepackage{mathrsfs}%
\usepackage[title]{appendix}%
\usepackage{xcolor}%
\usepackage{textcomp}%
\usepackage{manyfoot}%
\usepackage{booktabs}%
\usepackage{algorithm}%
\usepackage{algorithmicx}%
\usepackage{algpseudocode}%
\usepackage{listings}%
\usepackage{subfig}

\theoremstyle{thmstyleone}%
\theoremstyle{thmstyletwo}%

\theoremstyle{thmstylethree}%

\begin{document}

\title[Article Title]{High Power RF Pulse Shaping Tests with NG-LLRF and Cool Copper Collider Prototype Structure}


\author*[1]{\fnm{Chao} \sur{Liu}}\email{chaoliu@slac.stanford.edu}

\author[1]{\fnm{Ankur} \sur{Dhar}}
\author[1]{\fnm{Dennis} \sur{Palmer}}
\author[2]{\fnm{Ronald} \sur{Agustsson}}
\author[2]{\fnm{Diego} \sur{Amirari}}
\author[1]{\fnm{Martin} \sur{Breidenbach}}
\author[1]{\fnm{Emilio} \sur{Nanni}}

\affil*[1]{ \orgname{SLAC National Accelerator Laboratory}, \orgaddress{\street{2575 Sand Hill Road}, \city{Menlo Park}, \postcode{94025}, \state{California}, \country{USA}}}

\affil[2]{\orgname{RadiaBeam Technologies LLC}, \orgaddress{\street{1717 Stewart St}, \city{Santa Monica}, \postcode{90404}, \state{California}, \country{USA}}}


\abstract{RF pulse modulation techniques are widely applied to shape RF pulses for various types of RF stations of particle accelerators. The amplitude and phase modulations are typically implemented with additional RF components that require drive or control electronics. For the RF system-on-chip (RFSoC) based next generation LLRF (NG-LLRF) platform, which we have developed in the last several years, RF modulation and demodulation are fully implemented in the digital domain. Therefore, arbitrary RF pulse shaping can be realized without any additional analogue components. We performed a range of high-power experiments with the NG-LLRF and a prototype Cool Copper Collider (C\(^3\)) structure. In this paper, the RF field measured at different stages with different pulse shapes and peak power levels up to 5.4 MW and pulse width about 1 microsecond will be demonstrated and analyzed. The high precision pulse shaping schemes of the NG-LLRF can be applied to realize the phase modulation for a linear accelerator injector, the phase reversal for a pulse compressor, or the modulation required to compensate for the beam loading effect.}




\maketitle

\section{Introduction}\label{sec1}

The low-level radio frequency (LLRF) systems for particle accelerators typically stabilize the electromagnetic field in accelerator cavities with extremely high precision to achieve the optimum beam quality. As accelerator technology advances, the functional requirements for the LLRF have been extended. From single bunch per RF pulse to multiple bunch, intra-pulse stability control will be required and on top of inter-pulse control. More aggressively, new accelerator concepts, such as programmable accelerators, have been proposed that will require the LLRF to deliver RF pulses with an arbitrary envelope with extremely high precision. With the programmable accelerator, the charge, beam energy, repetition rate, and other operation parameters can be rapidly tuned in real-time. The enormous flexibility of operation can enable for more sustainable, more adaptable, and potentially more ground breaking high-energy physics experiments.

The direct RF sampling technique can significantly simplify the analogue RF circuit compared with that of heterodyne-based system architecture, but the RF performance of the sampling technique was the main concern for our targeted applications.  Since RFSoC has been released in 2018, we have investigated the continuous-wave (CW) RF performance of the RFSoC data converters in different frequency ranges and developed receiver and readout platforms for various physics experiments \cite{liu2021characterizing, liu2022development,  liu2023evaluating, henderson2022advanced, liu2023higher,liu2024development}. The RFSoC demonstrated high RF stability in CW test setups\cite{liu2021characterizing, liu2023evaluating}, but a range of particle accelerators operate with RF pulses. Therefore, we have conducted a variety of pulse-to-pulse stability studies for C-band \cite{liu2024direct, liu2024next,liu2025compact,liu2025high,liu2025nextical} and S-band LLRF systems for accelerators \cite{liu2025sband}. We have designed the first LLRF prototype for our initial target application, the new proposed lepton collider Higgs factory, Cool Copper Collider (C\(^3\)). The prototype demonstrated around 80 femtoseconds (fs) \cite{liu2025high} phase jitter with a direct loopback and the 150 fs requirement of C\(^3\) \cite{nanni2023status} is highly achievable with the NG-LLRF platform. In \cite{c3_2025esppu}, NG-LLRF has been identified as the LLRF solution for C\(^3\). In this paper, we will present a feasibility study of scaling up the NG-LLRF for the full C\(^3\) from architectural and cost perspectives. 

As introduced in \cite{liu2025high}, we performed preliminary high-power tests  with C\(^3\) accelerating structure cavities using the test facility at Radiabeam. The prototype C\(^3\) structure was also driven by RF pulses with different modulation schemes with potential applications at different types of accelerator RF station. In \cite{liu2025pulse}, we discuss a selection of preliminary test results with different pulse schemes at peak power level of around 16.45 MW and pulse width less than 500 ns.  In this paper, the test results with pulse width around  1 \(\mu\)s, which is long enough to show the full field filling process.  The pulses captured with square modulation and alternative modulations will be compared to demonstrate the flexibility of the NG-LLRF in arbitrary pulse shaping and measurement.

\section{High-Power Test of NG-LLRF with Cool Copper Collider Prototype Accelerating Structure with Different RF Pulse Shapes}

\subsection{High-power Test Setup}

The high-power test was performed with a high-power C-band  test stand at Radiabeam, which was described in detail in \cite{liu2025high}. RF pulses are synthesized directly by an integrated DAC in the RFSoC device of the NG-LLRF. A baseband pulse is loaded into the FPGA from the server via the GbE link. Then the baseband pulse is interpolated and up converted with a digital mixer. The DAC decodes the digital samples and is synthesized by an integrated DAC. In this test, the NG-LLRF generates the RF pulse with frequency around 5.712 GHz and the RF pulse feeds to a solid state amplifier (SSA) to drive the klystron. RF power is injected to the C\(^3\) prototype structure. There are couplers at three stages of the test setup, the klystron forward, the cavity forward, and cavity reflection. The RF signals from the couplers are attenuated and looped back to NG-LLRF. The three RF signals are directly sampled by the integrated ADCs in the fifth-order Nyquist. The digital samples are down converted with digital mixer at the corresponding RF frequency and then filtered and decimated. The baseband pulses of the three RF channels are recorded for further analysis.

The rated peak power of the test stand is approximately 25 MW with a pulse width of several \(\mu\)s, but this test was performed before the structure was fully conditioned. In this test, we have performed tests with two test configurations, 1 \(\mu\)s pulse width at 10 Hz at 5.2 MW peak power and 500 ns pulse width at 60 Hz at 16.45 MW peak power. 

\subsection{RF Pulse with a Linear Phase Ramp}

\begin{figure}[!tbp]
  \centering
  \subfloat[The forward power measured.]{\includegraphics[width=0.5\textwidth]{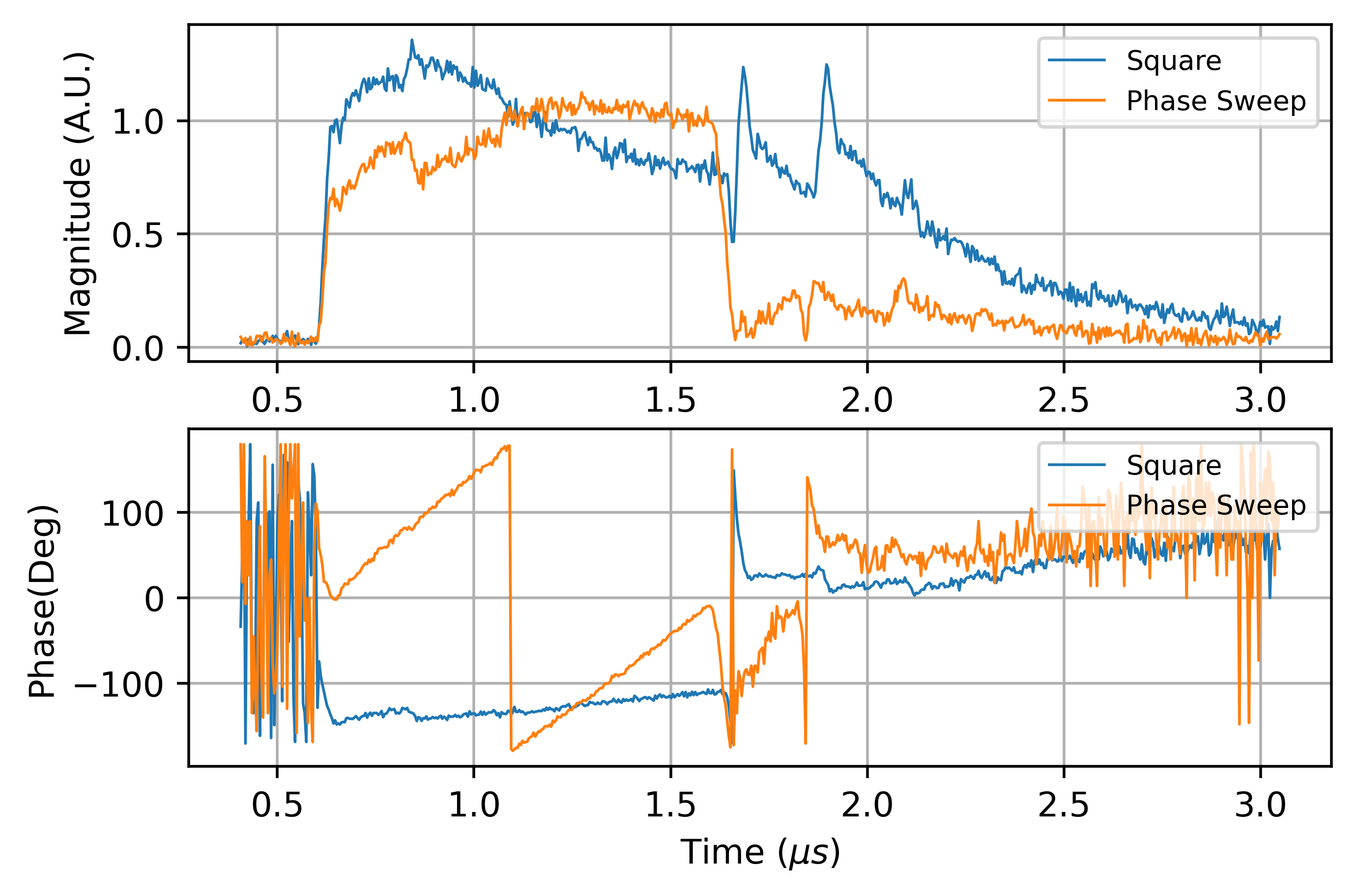}\label{fig:f4a}}
  \hfill
  \subfloat[The reflected power measured.]{\includegraphics[width=0.5\textwidth]{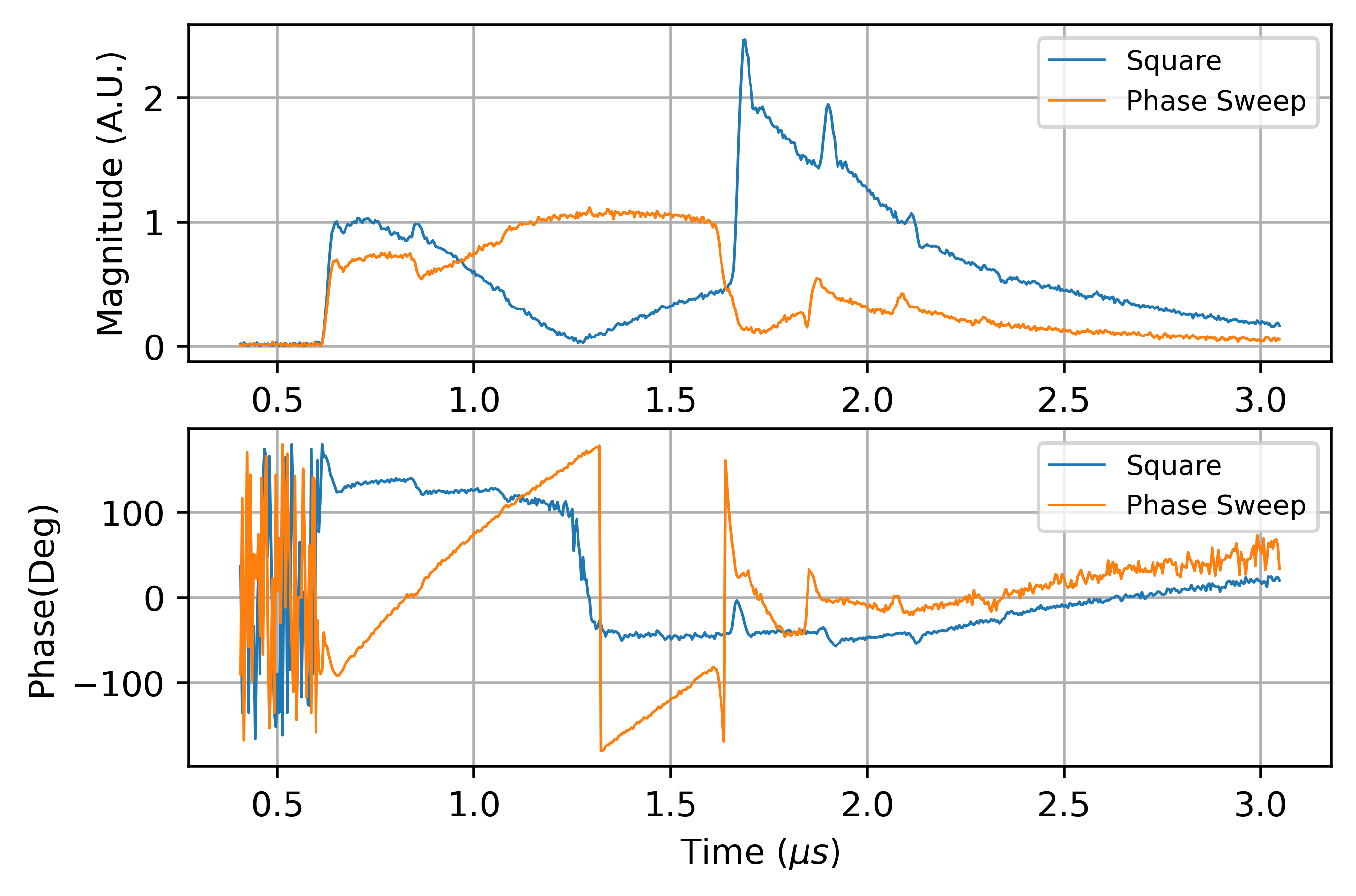}\label{fig:f4b}}
  \caption{The magnitude and phase of baseband pulses from high power test stand driven by RF pulses modulated with square and linear phase ramp envelope. The pulse width is about 1 \(\mu\)s and the peak power injected to the prototype structure is around 5.2 MW.}
\end{figure}

We performed high-power tests with different modulation schemes to evaluate the ability of NG-LLRF to synthesize and measure RF pulses with a variety of pulse shapes. In this test, we used a baseband pulse with a 360 degree linear phase ramp and fixed magnitude  modulation over the pulse duration. The peak power injected into the structure for both types of RF pulses is 5.2 MW peak power, 1 \(\mu\)s pulse width and 10 Hz repetition rate. In this case, the duration of the 360\textdegree \(\) linear phase ramp has been extended to 1 \(\mu\)s, which can be considered to drive the prototype cavity off resonance by 1 MHz. Figure \ref{fig:f4a} and \ref{fig:f4b} show the forward and reflection baseband signals for the square and phase ramp modulation schemes. With this linear phase ramp, the power injected into the structure is also almost fully reflected, since the bandwidth of the cavity structure is on the order of kilohertz. With a 1 \(\mu\)s square pulse, the structure is fully filled in about 700 ns, since the magnitude of the cavity reflection reached zero around that time. After the cavity is fully filled, the magnitude of the reflected signal increases linearly and the phase reversed, as the forward power cannot be filled into the cavity any more and got fully reflected. When the RF is switched off, there is a spike in the reflected signal, which is the desired behavior of an over coupled cavity design, and there is no beam presented in this case.

This test demonstrated that the NG-LLRF is capable of synthesizing and measuring the RF pulses with high precision phase modulation schemes in the C-band without any analog up and down conversions. With flexible phase modulation, phase stabilization within each RF pulse, beam load effect compensation on phase, or other phase modulation schemes that can improve the operation of the accelerating structures or other types of high-power RF stations can be implemented without substantial engineering efforts.

\subsection{RF Pulse with Phase Reversals}

\begin{figure}[!tbp]
  \centering
  \subfloat[The forward power measured.]{\includegraphics[width=0.5\textwidth]{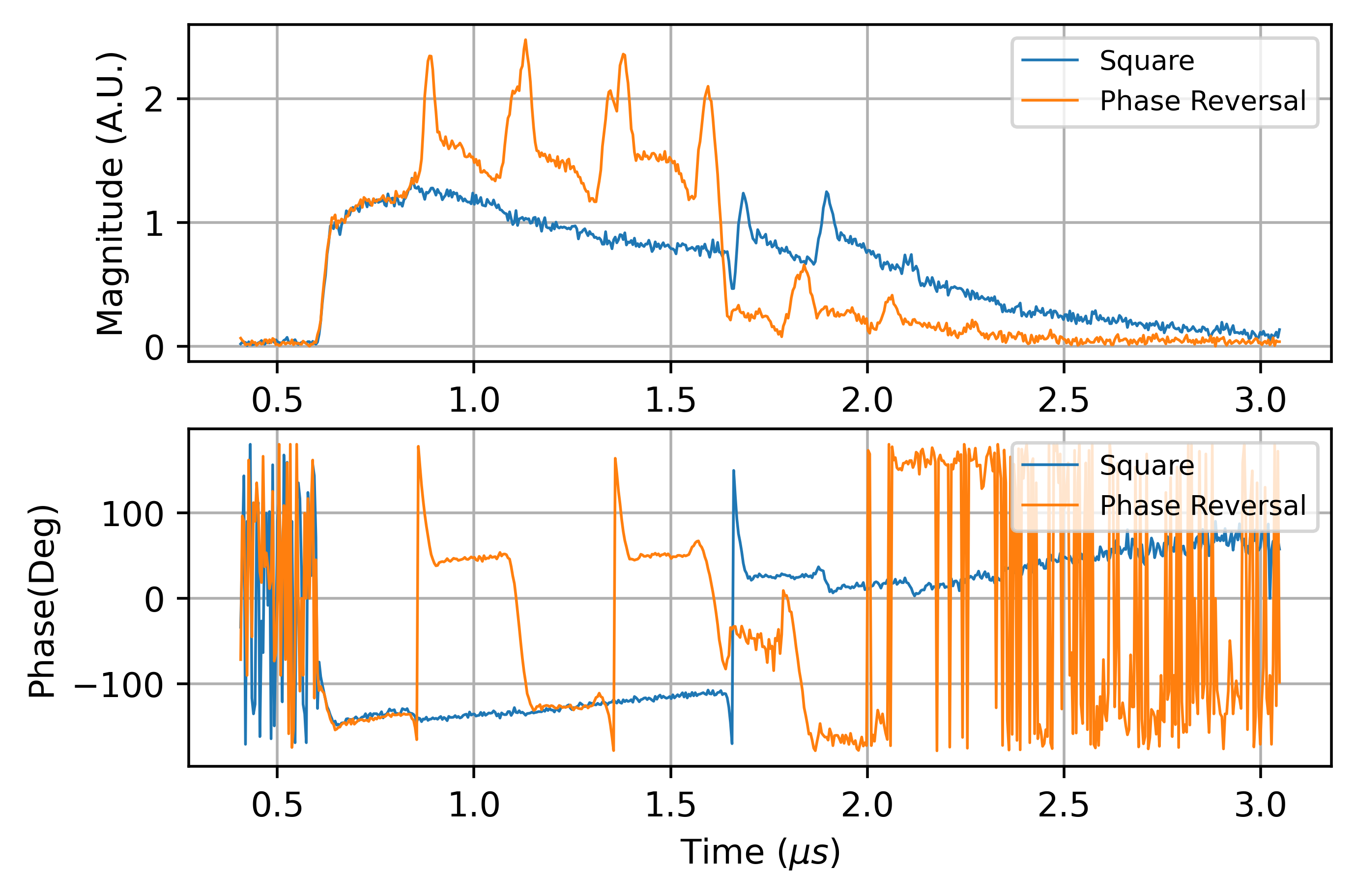}\label{fig:f6a}}
  \hfill
  \subfloat[The reflected power measured.]{\includegraphics[width=0.5\textwidth]{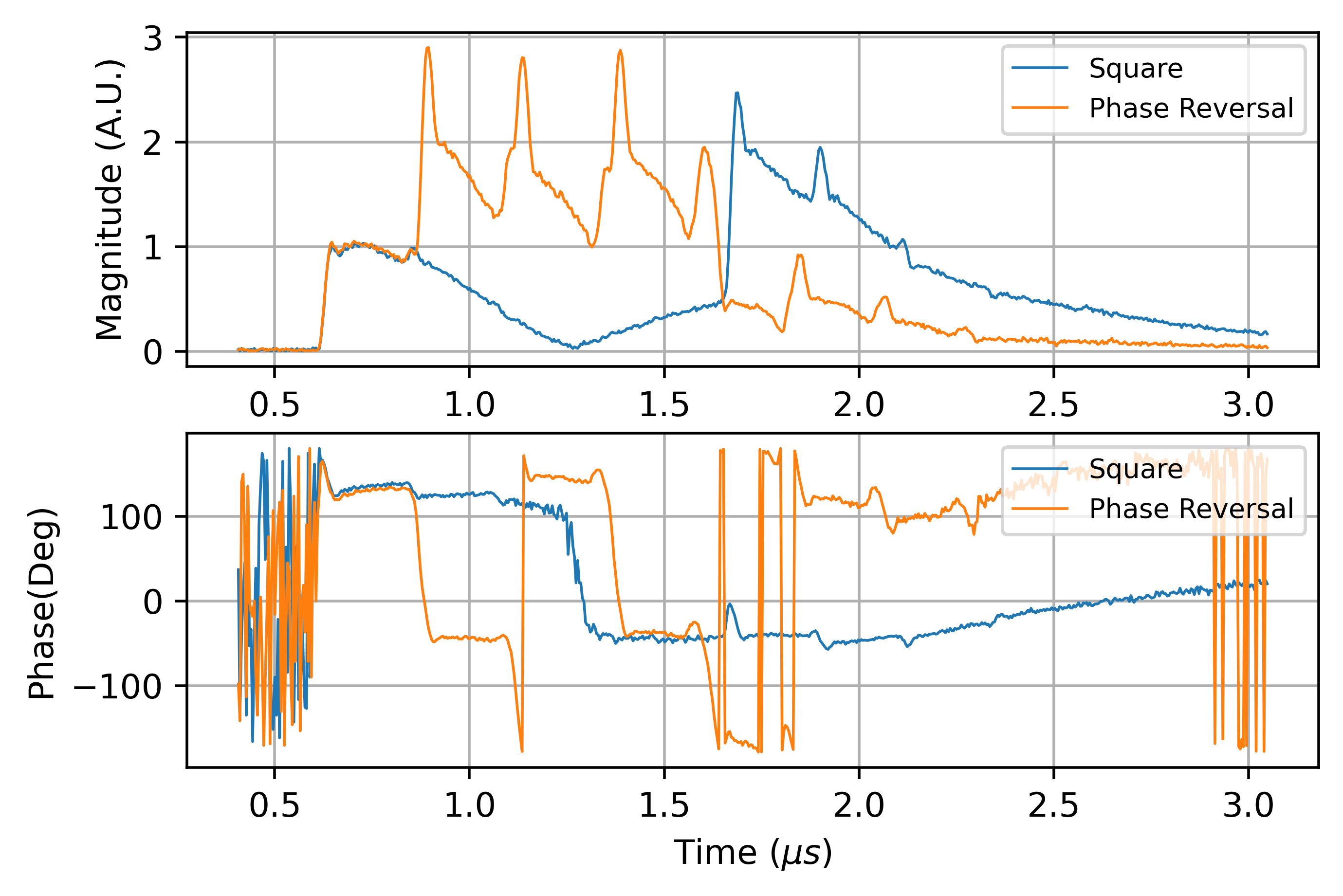}\label{fig:f6b}}
  \caption{The magnitude and phase of baseband pulses from high power test stand driven by RF pulses modulated with square and square wave with phase reversal every 250 ns envelopes. The pulse width is about 1 \(\mu\)s and the peak power injected to the prototype structure is around 5.2 MW.}
\end{figure}

Phase reversal is a common phase modulation scheme for SLAC Energy Doubler (SLED) type pulse compressors \cite{farkas1974sled}. The prototype C\(^3\) structure is not designed for pulse compression, but was tested with RF pulses with phase reversal to demonstrate the ability of NG-LLRF to generate phase flip. Figure \ref{fig:f5a} shows the forward coupler measurements with square and phase reversal RF drive modulation schemes with peak power at 5.2 MW and we can realize 3 flip events in each RF pulse. The cavity forward signal shown in Figure \ref{fig:f6a} shows that the 3 phase flips were successfully generated with the RF network. Figure \ref{fig:f6b} shows the reflected signals with two types of drive. The initial phase flip shows the SLED like rapid power extraction and the second flip shows the extraction in reversed phase again. For an SLED-based pulse compressor, the long pulse tail after the desired short pulse can potentially cause breakdowns and affect the peak power that could be delivered. Therefore, another  phase flip that extracts power rapidly and shortens the pulse tail can help reduce the breakdown rates. That has been demonstrated with the second phase flip in this test. The pulse width can be optimized in future design. This test demonstrates the flexibility of the NG-LLRF platform in phase modulation and could be the enabling technique for implementing an SLED-based pulse compressor with even high peak power in the future.

\subsection{Pulse Train within each RF Pulse}

The RF drives with phase modulation schemes were demonstrated in the previous two sections. The test results with an RF drive that has an amplitude modulation, an evenly spaced pulse train, will be described.

\begin{figure}[!tbp]
  \centering
  \subfloat[The forward power measured.]{\includegraphics[width=0.5\textwidth]{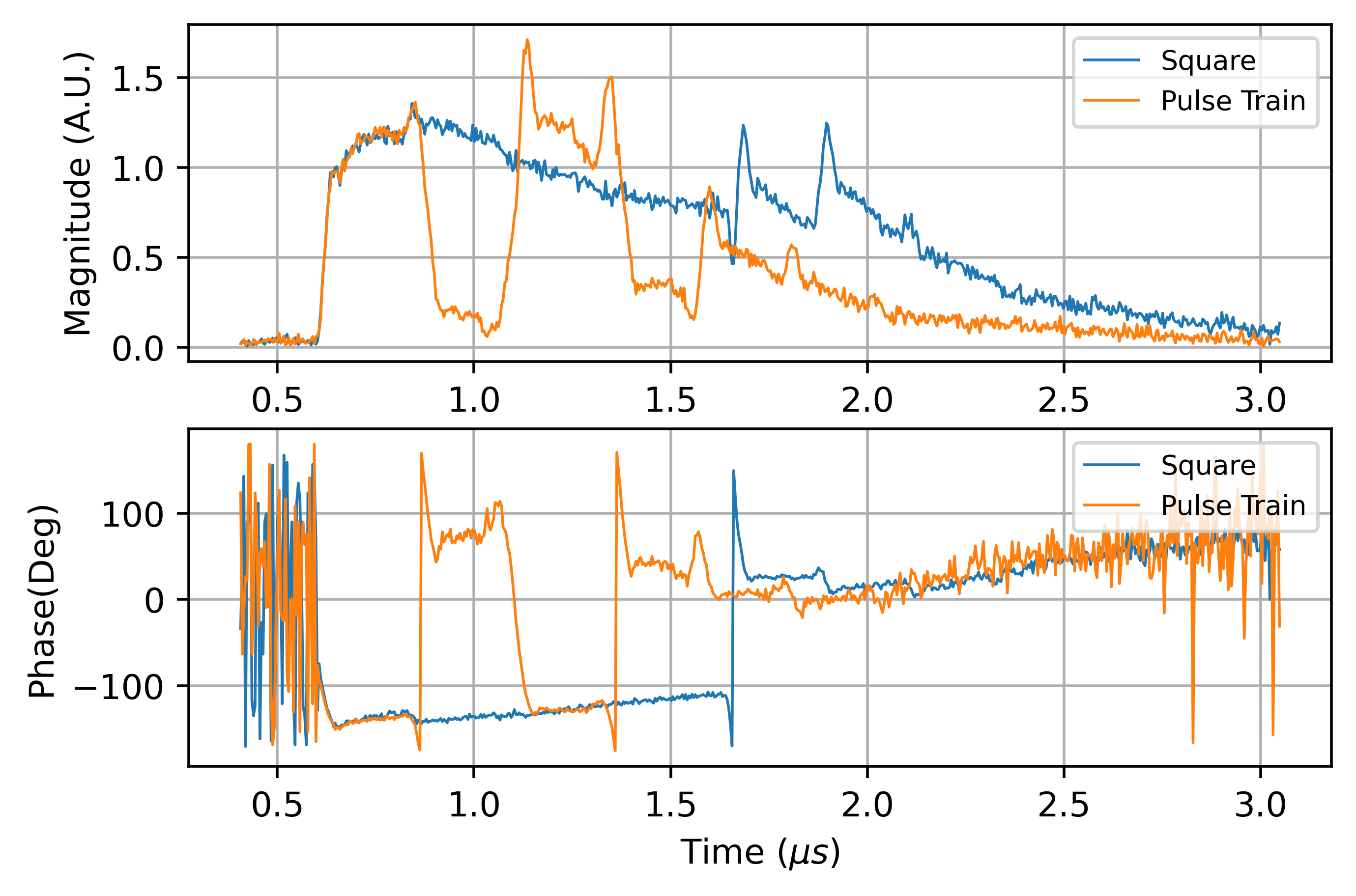}\label{fig:f8a}}
  \hfill
  \subfloat[The reflected power measured.]{\includegraphics[width=0.5\textwidth]{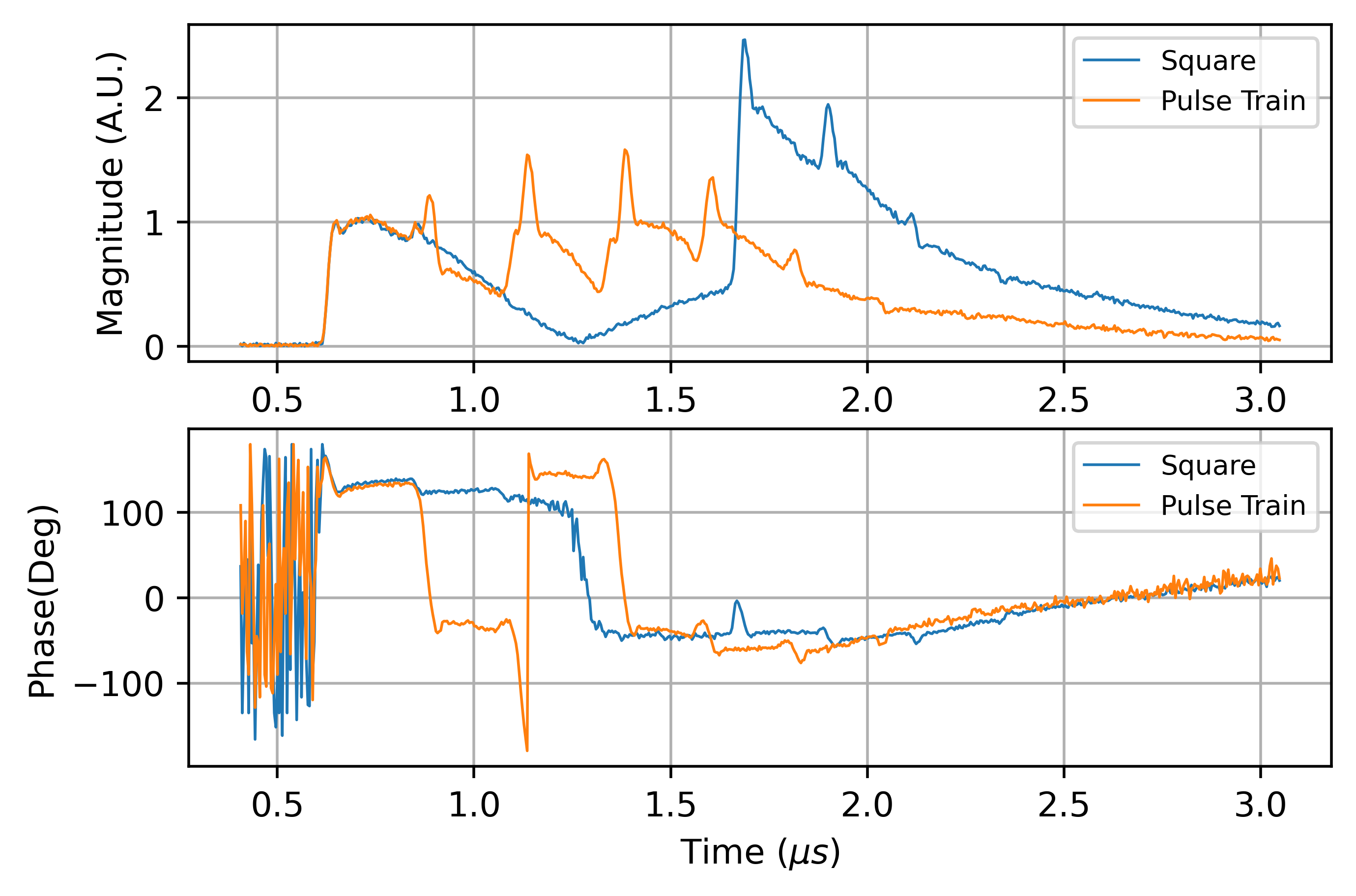}\label{fig:f8b}}
  \caption{The magnitude and phase of baseband pulses from high power test stand driven by RF pulses modulated with square and pulse train switch on and off every 250 ns envelopes. The pulse width is about 500 ns and the peak power injected to the prototype structure is around 5.2 MW.}
\end{figure}

Figure \ref{fig:f8a} shows the forward coupler measurements with square and pulse train modulation. When the RF is switched off after the initial 250 ns, the magnitude of the forward RF measurement does not drop to close to zero as expected. This is the result of cross coupling from reflection to forward, which is discussed in \cite{liu2025high}. However, we can see that the amplitude modulation was successfully applied. When the RF is switched off, the peak in the cavity reflection signal with pulse train modulation is lower than that with square modulation, since no RF power filled the structure after 250 ns with pulse train modulation.

The test was also performed with 1 \(\mu\)s RF pulses with pulse train modulation at peak power of 5.2 MW.  The cavity forward measurements shown in Figure \ref{fig:f8a} demonstrate that the RF pulse train was successfully generated and delivered to the prototype structure. Figure \ref{fig:f8b} cavity reflection measurements with the RF drive with two modulation schemes. The field fills and dissipates with the pulse train drive, and the RF power flow can be monitored from the reflection signal. 

The pulse train modulation tested in this case is to demonstrate flexible amplitude modulation of the high power RF with an NG-LLRF system. Amplitude modulation can be used for many stations in LINACs. For accelerators operating with bunch trains, the RF field in the structure within is required to be consistent with in each pulse. The RF network and beam loading often introduced uniform distortions to the pulses, and the shape of RF pulses generated by the NG-LLRF need to be tuned to compensate for such effects. The amplitude may also be used for electron bunch generation, such as defining the micro-pulse features and carving out macro-pulses at a variable width.

\section{Conclusion}

In this paper, we have summarized a selection of test results from a high-power test with a prototype NG-LLRF  and a prototype C\(^3\) accelerating structure focusing on the RF responses with different modulation schemes that are meaningful for future applications. For the direct RF synthesizing and sampling technique we adapted for NG-LLRF, the modulation and demonstration are performed fully in the digital domain. Therefore, we can load an arbitrary waveform to the FPGA fabric and get it modulated. Then the integrated DAC decodes the digital samples and synthesizes the RF pulses directly. The integrated ADCs sample the RF signals directly and down-converted the signal to baseband in the digital domain too. 

We have tested the system with the RF drive modulated with a linear phase ramp. The test demonstrated the high precision phase modulation capability of the NG-LLRF and the behavior of the prototype C\(^3\) structure with an off-resonance drive. Phase flip is also tested with the high-power test stand, as it is an important technique used for SLED-based pulse compressor or other rapid field power extraction applications. The test results demonstrated the rapid power extraction on the reflection signal, which is exactly expected for an SLED structure. The phase flip time with the C-band high power test network measure in this case is below 4 ns, which could be used as a baseline for future C-band pulse compressor design. The bandwidth of the measurement of NG-LLRF is 245.76 MHz, so the resolution of time is about 4 ns. The actual flip time might be even shorter. The system was also tested with an RF drive that has a pulse train pattern. The test results successfully showcased pulse train modulation, which can be adapted for more complex pulse train generation for electron injectors. The modulation schemes tested in this case cover only several commonly used cases and could be extended to arbitrary RF pulse shapes depending on the applications. The high flexibility and high precision RF shaping of the NG-LLRF could enable the implementation of the programmable accelerator concept.

\section{Acknowledgment}

The work of the authors is supported by the U.S. Department of Energy under Contract No. DE-AC02-76SF00515.

\section{Data Availability}
The data underlying this article will be shared on reasonable request to the corresponding author.

\bibliography{sn-bibliography}

\end{document}